\documentstyle[12pt]{article}
\begin{document}

\begin{flushright}
       {\bf UK/94-03}  \\
       Oct. 1994      \\
      hep-lat/9411067
\end{flushright}
\begin{center}

{\bf {\LARGE  Quark Model From Lattice QCD}}

\vspace{1.5cm}

{\bf  Keh-Fei Liu
\footnote{Talk presented at Int. Conf. High Energy Phys., Galsgow,
 July, 1994} and Shao-Jing Dong} \\ [0.5cm]
{\it  Dept. of Physics and Astronomy  \\
  Univ. of Kentucky, Lexington, KY 40506}
 \\[1em]

\end{center}

\vspace{1cm}

\begin{abstract}

We study the valence approximation in lattice QCD
of hadrons where the cloud quarks and antiquarks are
deleted by truncating the backward time propagation (Z graphs)
in the connected insertions. Whereas,
the sea quarks are eliminated via the quenched
approximation and in the disconnected insertions.
It is shown that the ratios of isovector to isoscalar matrix
elements in the nucleon reproduce the SU(6) quark model
predictions in a lattice QCD calculation. We also discuss
how the hadron masses are affected.

\end{abstract}

\section{Introduction}

In addition to its classification scheme, the quark model is,
by and large, quite successful in delineating the
spectrum and structure of mesons and baryons. One
often wonders what the nature of the approximation is, especially
in view of the advent of quantum chromodynamics (QCD).
In order to answer this question, we need to understand first where
the quark model is successful and where it fails.

To begin with, we need to define what we mean by the
quark model. We consider the simplest approach which includes the
following ingredients:
\begin{itemize}
\item
The Fock space is restricted to the valence quarks only.
\item
 These valence quarks,
be them the dressed constituent quarks or the bare quarks, are
confined in a potential or a bag. To this zeroth order, the hadron
wavefunctions involving u,d, and s quarks
are classified by the totally symmetric wavefunctions
in the flavor-spin and orbital space according to the $SU(6) \times
O(3)$ and totally antisymmetric/symmetric in the color space for the
baryons/mesons.
\item
The degeneracy within the the multiplts are lifted by the different
quark masses and the
residual interaction between the quarks which is
weak compared to the confining potential.
The one-gluon exchange potential is usually taken as this residual
interaction to describe the
hyper-fine and fine splittings of the hadron masses.
\end{itemize}


Given what we mean by the quark model, it is easier to understand where
the quark model succeeds and fails. It is successful in describing
hadron masses, relations of coupling and decay
constants, magnetic moments, Okubo-Zweig rule, etc. It is worthwhile
noting that all these are based on the valence picture aided with
$SU(6) \times O(3)$ group for its color-spin and space group. On the
other hand, it fails to account for the U(1) anomaly (the $\eta'$ mass)
, the proton spin crisis and the $\pi N \sigma$ term. All these
problems involve large contribution from disconnected insertions
involving sea-quarks~\cite{liu92}. It is natural not to expect the
valence quark model to work. There are other places where the
valence quark model does not work well. These include $\pi\pi$,
$\pi N$ scatterings, current algebra relations, and the form factors
of the nucleon which are better described by meson effective
theories with chiral symmetry taken into account. For example, the
$\pi\pi$ scattering is well described in the chiral perturbation
theory, the $\pi N$ scattering and the nucleon electromagnetic
,axial, and pseudoscalar form factors
(especially the neutron charge radius), Goldberg-Treiman relation
are all quite well given in the skyrmion approach~\cite{ll87}.
One common theme
of these models is the chiral symmetry which involves meson cloud
and hence the higher Fock space beyond the valence.

\section{Valence Approximation}

It is then clear that there are three ingredients in the classification
the quarks, i.e. the valence, the cloud, and the sea quarks.
The question is how one defines them unambiguously and
in a model independent way in QCD. It has been shown recently
\cite{ld94} that in evaluating the hadronic tensor in the
deep inelastic scattering, the three topological distinct
contractions of quark fields lead to the three quark-line skeleton
diagrams. The self-contraction of the current leading to
a quark loop is separated from the quark lines joining
the nucleon interpolating fields. This disconnected insertion (D.I.)
refers to the quark lines which are of courses connected by
the gloun lines. This D.I.
defines the sea-parton. One class of the connected
insertion (C.I.) involves
an anti-quark propagating backwards in time between the currents and
is defined as the cloud anti-quark. Another class of the C.I. involves
a quark propagating forward in time between the currents and is
defined to be the sum of the valence and cloud quarks.
Thus, in the parton
model, the antiquark distribution should be written as
\begin{equation}   \label{antiquark}
\overline{q}^i(x) = \overline{q}_c^i(x) + \overline{q}_s^i(x).
\end{equation}
to denote their respective origins for
each flavor i. Similarly, the quark distribution is written as
\begin{equation}   \label{quark}
q^i(x) = q_V^i(x) + q_c^i(x) + q_s^i(x)
\end{equation}
Since $q_s^i(x) = \overline{q}_s^i(x)$,
we define $q_c^i(x) = \overline{q}_c^i(x)$ so that
$q_V^i(x)$ will be responsible for the baryon number,
i.e. $\int u_V(x) dx = \int [u(x) - \overline{u}(x)] dx = 2$ and
$\int d_V(x) dx = \int [d(x) - \overline{d}(x)] = 1$ for the proton.

We can reveal the role of these quarks in the nucleon matrix elements
which involve the three-point function with one current.
The D.I. in the three-point function involves the
sea-quark contribution to the m.e. It has been shown that the
this diagram has indeed large contributions for the flavor-singlet
scalar and axial charges \cite{dl93} so that the discrepancy between
the valence quark model and the experiment in the $\pi N \sigma$ term
and the flavor-singlet $g_A$ can be understood. Thus we conclude
that in order to simulate the valence quark model, the first step
is to eliminate the quark loops. This can be done in the well-known
quenched approximation by setting the fermion determinant to a
constant.

In order to reveal the effect of the cloud
degree of freedom, we have calculated the ratios of the
isoscalar to isovector axial and scalar charges in a quenched lattice
calculation. The ratio of the isoscalar (the C.I. part) to isovector
axial charge can be written as

\begin{equation} \label{axial}
R_A = \frac{\langle p|\bar{u}\gamma_3\gamma_5 u
+ \bar{d}\gamma_3\gamma_5 d|p\rangle}
{\langle p|\bar{u}\gamma_3\gamma_5 u  -
\bar{d}\gamma_3\gamma_5 d|p\rangle}\, \begin{array}{|l}
   \\ \footnotesize{C.I.} \end{array} \!\!
 =\, \frac{g_A^1}{g_A^3} \,\begin{array}{|l}
 \footnotesize{C.I.} \end{array}  \!\!
 =\, \frac{\int dx [\Delta u(x)
+ \Delta d(x)]} {\int dx [\Delta u(x) - \Delta d(x)]} \,
\begin{array}{|l}  \\ \footnotesize{C.I.} \end{array}
\end{equation}
where $\Delta u(\Delta d)$ is the polarized parton distribution
of the u(d) quark and antiquark in the C.I.
 For the non-relativistic case, $g_A^3$
is 5/3 and $g_A^1$ for the C.I. is 1
Thus, the ratio $R_A$ should be 3/5. Our lattice results
based on quenched $16^3 \times 24$ lattices with $\beta = 6$ for
the Wilson $\kappa$ ranging between 0.154 to 0.105 which correspond
to strange and twice the charm masses are plotted in Fig. 1 as a
function of the quark mass $ma = ln (4\kappa_c/\kappa -3)$.
We indeed find this ratio for the heavy quarks (i.e. $\kappa \geq
0.133$ or $ma \geq 0.4$ in Fig.1). This is to
be expected because the cloud antiquarks which involves Z-graphs
are suppressed for non-relativistic quarks by $O(p/m_q)$.
Interestingly, the ratio dips under 3/5 for light quarks.
We interpret this as
due to the cloud quark and antiquark, since in the relativistic
valence quark models (i.e. no cloud nor sea quarks)
the ratio remains to be 3/5. To verify that this is indeed
caused by the cloud antiquarks from the backward time propagation,
 we perform the following approximation.
In the Wilson lattice action, the backward time hopping
is prescribed by the term \mbox{$- \kappa (1 - \gamma_4) U_4
(x) \delta_{x,y-a_4}$}. We shall amputate this term from the
quark matrix in our calculation of the quark propagators.
As a result,
the quarks are limited to propagating forward in time and
there will be no Z-graph and hence no cloud quarks and antiquarks. The
Fock space is limited to 3 valence quarks. Thus we shall refer to
this as the {\it valence approximation} and we believe it simulates
what the naive quark model is supposed to describe by design.
After making this valence approximation for the light quarks
with $\kappa = 0.148, 0.152,$ and 0.154 (The quark mass turns out
to differ from before only at the perturbative one-loop order,i.e.
$O(\alpha_s)$, which is very small.
we find that the ratio $R_A$ becomes 3/5 with errors less than the
size of the circles in Fig. 1. Since the valence quark model prediction
of $R_A$ is well reproduced by the valence approximation, we believe
this proves our point that the deviation of $R_A$ from 3/5 in Fig. 1
is caused by the backward time propagation, i.e. the cloud quarks
and antiquarks.

Similar situation happens in the scalar matrix elements.
In the parton model description of the forward m.e., the ratio of the
isovector to isoscalar scalar charge
of the proton for the C.I. is then approximated
according to eqs. (\ref{antiquark}) and (\ref{quark}) as
\begin{equation}   \label{scalar}
R_S= \frac{\langle p|\bar{u}u  - \bar{d}d|p\rangle}
{\langle p|\bar{u}u  + \bar{d}d|p\rangle}
\, \begin{array}{|l}  \\ \footnotesize{C.I.} \end{array} \!\!
=\,\frac{1 + 2\int dx [\bar{u}_c(x) - \bar{d}_c(x)]}
{3 + 2\int dx [\bar{u}_c(x) + \bar{d}_c(x)]}
\end{equation}
Since the quark/antiquark number is positive definite,
we expect this ratio to be $ \leq 1/3$. For
heavy quarks where the cloud antiquarks are suppressed,
the ratio is indeed 1/3 (see Fig. 2). For quarks
lighter than $\kappa = 0.140$, we find that the ratio is in fact less
than 1/3.
The lattice results of the valence approximation for the light
quarks, shown as the circles in Fig. 2,
turn out to be 1/3. This shows that the deviation
of $R_S$ from 1/3 is caused by the cloud quarks and antiquarks.
With these findings, we obtain an upper-bound for the
violation of GSR~\cite{ld94}, i.e.
$n_{\bar{u}} - n_{\bar{d}} \leq -0.12 \pm 0.05$. This clearly shows
that $n_{\bar{u}}- n_{\bar{d}}$ is negative and is quite consistent
with the experimental result $\int dx [\bar{u}^p(x) - \bar{d}^p(x)]
= -0.14 \pm 0.024$.

\section{Hadron Spectroscopy}

To further explore the consequences of the valence approximation,
we calculate the baryon masses. Plotted in fig. 3 are masses of
$\Delta, N, \rho$, and $\pi$ as a function of the quark mass $ma$ on
our lattice with quenched approximation. We see that the hyper-fine
splittings between the $\Delta$ and N, and the $\rho$ and $\pi$
grow when the quark mass approaches the chiral limit as expected.
However, it is surprising to learn that in the valence approximation,
the $\Delta$ and N become degenerate within errors, so do the
$\rho$ and $\pi$ as shown in Fig. 4. Since the one-gluon exchange is
not switched off in the valence approximation, the hyper-fine
splitting is probably not due to the one-gluon exchange potential
as commonly believed. Since this is a direct consequence of
eliminating the cloud quark/antiquark degree of freedom, one can
speculate that it has something to do with the cloud. It seems
that a chiral soliton like the skyrmion might delineate a more accurate
dynamical picture than the one-gluon exchange spin-spin interaction.

To conclude, we find that the valence approximation in QCD reproduces
the SU(6) results of the valence quark model better than we
anticipated. Especially in hadron masses, the results seem to indicate
that there are no hyper-fine splittings, modulo the uncertainty
due to the statistical and systematic errors.

\noindent
{\bf Figure Captions:} \\
Fig. 1  The ratio $R_A$ of eq. (3) as a function of the
quark mass $ma = ln (4\kappa_c/\kappa -3)$. \\
Fig. 2  The ratio $R_S$ of eq. (4) as a function of the
quark mass ma.  \\
Fig. 3 Masses of $\Delta, N, \rho$, and $\pi$ (in lattice units)
as a function of the quark mass ma in the quenched approximation. \\
Fig. 4 The same as in Fig. 3 with the valence approximation.

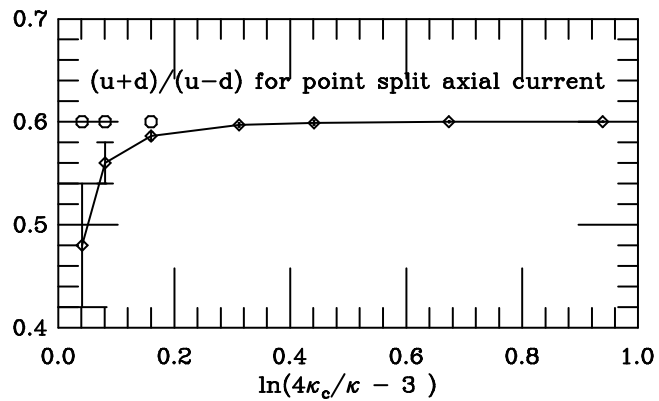
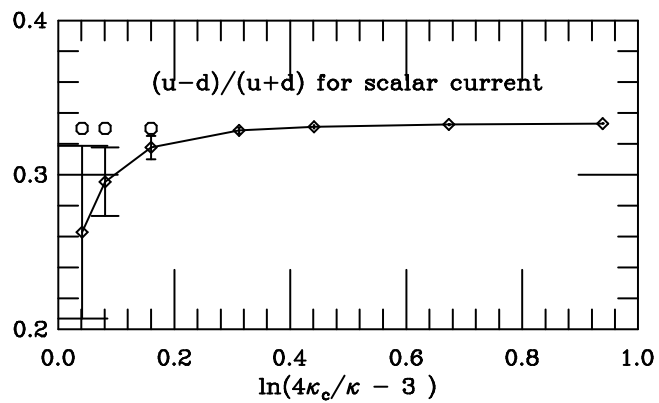
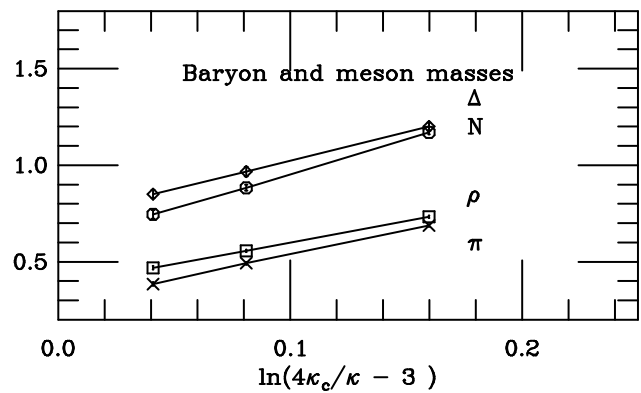
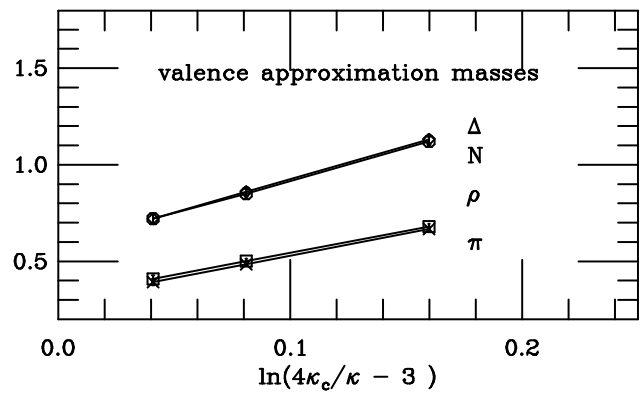
\end{document}